\documentclass[%
preprint,
showpacs,
pre,
amsmath,amssymb,
aps,
]{revtex4-1}

\usepackage{graphicx}
\usepackage{dcolumn}
\usepackage{bm}

\DeclareMathOperator{\Tr}{Tr}

\begin{document}

\title{Quantum annealing with antiferromagnetic fluctuations}

\author{Yuya Seki}
\author{Hidetoshi Nishimori}
\affiliation{Department of Physics, Tokyo Institute of Technology, Oh-okayama,
Meguro-ku, Tokyo 152-8551, Japan}

\date{\today}

\begin{abstract}
We introduce antiferromagnetic quantum fluctuations into
quantum annealing in addition to the conventional transverse-field term.
We apply this method to the infinite-range ferromagnetic $p$-spin model,
for which the conventional quantum annealing has been shown to have
difficulties to find the ground state efficiently
due to a first-order transition.
We study the phase diagram of this system
both analytically and numerically.
Using the static approximation, we find that there exists
a quantum path to reach the final ground state from the
trivial initial state that avoids first-order transitions
for intermediate values of $p$.
We also study numerically the energy gap
between the ground state and the first excited state
and find evidence for intermediate values of $p$ that
the time complexity scales polynomially with the system size
at a second-order transition point
along the quantum path that avoids first-order transitions.
These results suggest that quantum annealing would be able to solve
this problem with intermediate values of $p$ efficiently
in contrast to the case with only simple transverse-field fluctuations.
\end{abstract}


\maketitle

\section{\label{sec:intro}Introduction}
To find efficient algorithms for combinatorial optimization problems
is one of the important goals of computer science.
An algorithm is efficient if
its running time is bounded by a polynomial
in the problem size.
If an efficient algorithm for a problem is known,
the problem is considered as easy.
However, most of interesting combinatorial optimization problems
are hard, i.e., even the best known algorithms
take running times growing exponentially as the problem size increases~\cite{garey79:_comput_intrac, hartmann05:_phase_trans_combin_optim_probl}.

Many combinatorial optimization problems can be translated into
physics problems of finding the ground state (optimal solution)
of an Ising spin system~\cite{hartmann05:_phase_trans_combin_optim_probl,hopfield86:_comput_neural_circuit,fu86:_applic_statis_mechan_np_compl}.
The cost function corresponds to the energy of the system.
This transformation enables us to study combinatorial optimization problems
by ideas and methods developed in physics.
An interesting example is quantum annealing.

Quantum annealing~\cite{kadowaki98:_quant_anneal_trans_ising_model, finnila94:_quant_anneal, das08:_collo_quant_anneal_quant_compu}
(and its cousin, quantum adiabatic computation~\cite{farhi01:_quant_adiab_evolut_algor_applied})
is a quantum algorithm
to obtain an approximate solution for a combinatorial optimization problem,
which often outperforms simulated annealing~\cite{kirkpatrick83:_optim_simul_anneal,
morita08:_mathem_found_quant_anneal},
another
heuristic algorithm coming form physics.
Unlike simulated annealing,
QA uses tunneling effect caused by quantum fluctuations
to search for the optimal solution.
By controlling the strength of the fluctuations properly,
we can reach the solution with a high probability.

To be more explicit, let us consider the following
time-dependent Hamiltonian:
\begin{equation}
 \hat{H}(t) = s(t)\hat{H}_{0} + \bigl(1-s(t)\bigr)\hat{V},
\end{equation}
where $\hat{H}_{0}$ is the target Hamiltonian,
whose ground state is an optimal solution,
represented in terms of the $z$~components of the Pauli matrices
$\hat{\sigma }_{i}^{z}\ (i=1,\dotsc ,N)$.
The symbol $N$ denotes the number of spins.
The operator $\hat{V}$ is arbitrary as long as
it does not commute with $\hat{H}_{0}$ and has a unique trivial ground state.
This noncommutativity introduces quantum fluctuations into the system,
causing state transitions. It is thus called the driver Hamiltonian.
A typical example of the driver Hamiltonian is the transverse-field operator
$\hat{V}_{\text{TF}} \equiv -\sum _{i=1}^{N} \hat{\sigma }_{i}^{x}$,
where the $\hat{\sigma }_{i}^{x}\ (i=1,\dotsc ,N)$ are the $x$~components of
the Pauli matrix.
The control parameter $s(t)$ starts at zero $(s(0)=0)$
and increases monotonically to unity.
We assume that $s(\tau )=1$, i.e., the running time of QA is $\tau $.
For simplicity, the linear function $(s(t)=t/\tau )$ is adopted in most studies.
We then calculate the time evolution starting from the trivial ground state
$| \Psi (0)\rangle$ by solving the Schr\"{o}dinger equation,
\begin{equation}
 i\frac{d}{dt}| \Psi (t)\rangle = \hat{H}(t)| \Psi (t)\rangle ,\ 0\leq t \leq \tau .\label{eq:schrodinger equation}
\end{equation}
Note that $\hat{H}(0)=\hat{V}$ and $\hat{H}(\tau )=\hat{H}_{0}$,
and the initial state $| \Psi (0)\rangle$ is the ground state of $\hat{H}(0)$.
If we change the control parameter slowly ($\tau \gg 1$),
the state would stay very close to the instantaneous ground state
during the time evolution.
We can then achieve the ground state of $\hat{H}_{0}$ at $t=\tau $, which is
the optimal solution.

The condition to stay close to the ground state is expressed as
$\tau \gg \Delta _{\text{min}}^{\!-2}$ according to the adiabatic theorem~\cite{morita08:_mathem_found_quant_anneal},
where $\Delta _{\text{min}}$ is
the minimum energy gap from the ground state.
Thus the minimum gap determines the efficiency of QA for a given problem.
In the case that the minimum gap decays exponentially with the system size
as $\Delta _{\text{min}} \propto \exp (-\alpha N)$, the running time increases
exponentially. This means that QA cannot solve the problem efficiently.

J\"{o}rg \textit{et al.}\ have shown that
QA with the conventional transverse-field operator
costs exponentially long time
to reach the ground state of the ferromagnetic $p$-spin model for $p>2$~\cite{Jorg2010energy}.
The (target) Hamiltonian of this model is given by
\begin{equation}
 \hat{H}_{0} = -N\Bigl(\frac{1}{N}\sum _{i=1}^{N} \hat{\sigma }_{i}^{z}\Bigr) ^{p}.\label{eq:p spin model}
\end{equation}
In the $p\to \infty$ limit, this model reduces to the Grover problem~\cite{Jorg2010energy, grover97:_quant_mechan_helps_searc_needl_hayst},
which any known algorithms, even quantum algorithms, cannot solve efficiently.
J\"{o}rg \textit{et al.}\ have also shown that this system undergoes a quantum first-order phase transition
during the time evolution,
which is a characteristic feature of hard optimization problems~\cite{jorg08:_simpl_glass_model_their_quant_anneal, young10:_first_order_phase_trans_quant_adiab_algor, jorg10:_first_order_trans_perfor_quant}.

Nevertheless, the above result does not necessarily suggest a complete
failure of QA for this simple problem~(\ref{eq:p spin model}).
Note that the above result has been derived using
the transverse-field operator as a driver Hamiltonian.
This implies that different operators may lead to improved performance.
We show in this paper that an operator,
which induces antiferromagnetic fluctuations, significantly improves
the efficiency of QA for this model with intermediate values of $p$.

We organize the present paper as follows:
In Sec.~\ref{sec:AFF}, we define a quantum driver operator $\hat{V}_{\!\text{AFF}}$
and the total Hamiltonian $\hat{H}(t)$. We then
explain the idea of QA using antiferromagnetic fluctuations.
In Sec.~\ref{sec:analysis}, we calculate the partition function
with the static approximation and derive self-consistent equations.
From these equations, we analyze numerically  phase diagrams for finite $p$
in Sec.~\ref{sec:results}.
Numerical calculations show that
first-order transitions can be avoided if we change
the control parameters ingeniously.
In Sec.~\ref{sec:phase diagram for p infinity},
we discuss the $p\to \infty$ limit.
Finally, Sec.~\ref{sec:conclusion} is devoted to the conclusion.

\section{\label{sec:AFF}Antiferromagnetic fluctuations}
The main proposition of this paper is that an introduction of the
following antiferromagnetic interaction
\begin{equation}
 \hat{V}_{\!\text{AFF}} \equiv
  +N \Bigl(\frac{1}{N}\sum _{i=1}^{N} \hat{\sigma }_{i}^{x}\Bigr)^{2}
\label{eq:Vaff}
\end{equation}
in addition to the conventional transverse-field term $\hat{V}_{\text{TF}}$
greatly facilitates the process of quantum annealing
as shown in the following sections.
The total Hamiltonian is therefore
\begin{equation}
 \hat{H} (s,\lambda )
  = s\{ \lambda \hat{H}_{0} + (1-\lambda )\hat{V}_{\!\text{AFF}} \}
  + (1-s)\hat{V}_{\text{TF}},\label{hamiltonian s lambda}
\end{equation}
where the parameters $s$ and $\lambda $ should be changed appropriately
as functions of time.
The initial Hamiltonian has $s=0$ and $\lambda $ arbitrary,
and the final one has $s=\lambda =1$.
Intermediate values $(s,\lambda )$ should be chosen according to the
prescription given below.

This idea somewhat resembles that of quantum adiabatic algorithm with different paths~\cite{farhi2002:_quant_adia_evol_algo_with_diff_paths}.
This latter approach also considers a total Hamiltonian which consists of three parts:
a target, a driver, and another Hamiltonian $\hat{H}_{\text{E}}$
corresponding to $\hat{V}_{\!\text{AFF}}$.
Whereas $\hat{V}_{\!\text{AFF}}$ is defined uniquely,
the components of $\hat{H}_{\text{E}}$ are chosen randomly
according to some prescription.
For this system, one calculates repeatedly the time evolution
by changing $\hat{H}_{\text{E}}$.
Then, in some instances,
one reaches the optimal solution.
In contrast, our approach involves no stochastic processes
and therefore only a single run achieves the goal.

\section{\label{sec:analysis}Analysis with the static approximation}
We now confine ourselves to the ferromagnetic $p$-spin model with odd $p$
of Eq.~(\ref{eq:p spin model}) as the target Hamiltonian and analyze
the properties of $\hat{H}(s,\lambda )$.
This section is devoted to analytic computations.

\subsection{\label{subsec:partition func}Partition function}
We first calculate the partition function of the system
of Eq.~(\ref{hamiltonian s lambda}) at finite temperatures.
The partition function can be written in the following form
using the Suzuki-Trotter formula~\cite{suzuki76:_relat_dimen_quant_spin_system}:
\begin{align}
 Z &= \lim _{M\to \infty} Z_{M}
\notag \\
&\equiv  \lim _{M\to \infty }\Tr \bigl(e^{-\frac{\beta}{M} s\lambda \hat{H}_{0}}
  e^{-\frac{\beta}{M}\{ s(1-\lambda )\hat{V}_{\!\text{AFF}}
  + (1-s)\hat{V}_{\text{TF}}\}}\bigr)^{M}
\notag \\
&= \lim _{M\to \infty }\sum _{\{ \sigma _{i}^{z}\}} \langle \{\sigma _{i}^{z} \}|
\biggl(\exp \Bigl[\frac{\beta s \lambda N}{M}\Bigl(\frac{1}{N}\sum _{i=1}^{N}\hat{\sigma }_{i}^{z} \Bigr) ^{p}\Bigr]
\notag \\
&\hphantom{={}} \times
\exp \Bigl[-\frac{\beta s(1-\lambda )N}{M}\Bigl(\frac{1}{N}\sum _{i=1}^{N}\hat{\sigma }_{i}^{x}\Bigr)^{2}
+
\frac{\beta (1-s)}{M} \sum _{i=1}^{N}\hat{\sigma }_{i}^{x}
\Bigr] \biggr)^{M}| \{\sigma _{i}^{z}\}\rangle ,\label{eq:Z 1}
\end{align}
where $\sum _{\{\sigma _{i}^{z}\}}$ denotes the summation
over all spin configurations
in the $z$~basis,
and $| \{\sigma _{i}^{z}\} \rangle \equiv
\bigotimes _{i=1}^{N}| \sigma _{i}^{z}\rangle$.
The state $| \sigma _{i}^{z}\rangle$ is the eigenstate of $\hat{\sigma }_{i}^{z}$,
having the eigenvalue $\sigma _{i}^{z}$ $(=\pm 1)$.
Similar notations will be used for the $x$~basis.

We then introduce the following $M$ closure relations:
\begin{align}
  \hat{1}(\alpha ) &\equiv \sum _{\{\sigma _{i}^{z}(\alpha )\}}
  | \{\sigma _{i}^{z}(\alpha )\} \rangle
  \langle \{\sigma _{i}^{z}(\alpha )\}|
\notag \\
&\hphantom{={}}\times
\sum _{\{\sigma _{i}^{x}(\alpha )\}} |\{\sigma _{i}^{x}(\alpha )\}\rangle
 \langle \{\sigma _{i}^{x}(\alpha )\}| ,
\end{align}
where $\alpha =1,\dotsc ,M$.
Inserting $\hat{1}(\alpha )$ just before the $\alpha $th exponential operator
involving $\hat{\sigma }_{i}^{x}$ in Eq.~(\ref{eq:Z 1}),
we have
\begin{align}
& Z_{M} = \sum _{\{\sigma _{i}^{z}(\alpha )\}}
\sum _{\{\sigma _{i}^{x}(\alpha )\}}
\prod_{\alpha =1}^{M}
\exp \Bigl[
\frac{\beta s \lambda N}{M}\Bigl( \frac{1}{N}\sum _{i=1}^{N} \sigma _{i}^{z}(\alpha )\Bigr)^{p}
\notag \\ &
-\frac{\beta s(1- \lambda)N}{M}\Bigl( \frac{1}{N}\sum _{i=1}^{N}\sigma _{i}^{x}(\alpha )\Bigr)^{2}
+\frac{\beta (1-s )}{M}\sum _{i=1}^{N}\sigma _{i}^{x}(\alpha )
\Bigr]
\notag \\ &
\times \prod_{i=1}^{N}
\langle \sigma _{i}^{z}(\alpha ) | \sigma _{i}^{x}(\alpha ) \rangle
\langle \sigma _{i}^{x}(\alpha ) | \sigma _{i}^{z}(\alpha +1) \rangle
\end{align}
with periodic boundary conditions such that $\sigma _{i}^{z}(1)=\sigma _{i}^{z}(M+1)$
for $i=1,\dotsc ,N$.

To simplify the spin product terms $(\sum _{i=1}^{N}\sigma _{i}^{z}(\alpha )/N)^{p}$
and $(\sum _{i=1}^{N}\sigma _{i}^{x}(\alpha )/N)^{2}$,
we introduce the following
integral representation of the delta function:
\begin{align}
 \delta \Bigl(Nm - \sum _{i=1}^{N} \sigma _{i}\Bigr)
=\int d\tilde{m}
 \exp \Bigl[-\tilde{m}
 \Bigl(Nm - \sum _{i=1}^{N} \sigma _{i}\Bigr)\Bigr].
\label{delta function}
\end{align}
Here, $m$ denotes the magnetization (order parameter),
and its conjugate variable is $\tilde{m}$.
Using Eq.~(\ref{delta function}), we can rewrite $Z_{M}$ as
\begin{align}
 Z_{M} &= \sum _{\{\sigma _{i}^{z}(\alpha )\}}
\sum _{\{\sigma _{i}^{x}(\alpha )\}}
\prod_{\alpha =1}^{M}\idotsint
dm^{z}(\alpha )\,d\tilde{m}^{z}(\alpha )\,
dm^{x}(\alpha )\,d\tilde{m}^{x}(\alpha )
\notag \\
&\hphantom{={}}\times
 \exp \Bigl[N\Bigl(s\lambda \frac{\beta }{M}\bigl(m^{z}(\alpha )\bigr)^{p}
 -\tilde{m}^{z}(\alpha )m^{z}(\alpha )\Bigr)\Bigr]
\notag \\
&\hphantom{={}}\times
 \exp\Bigl[N\Bigl(-s(1-\lambda )\frac{\beta }{M}\bigl(m^{x}(\alpha )\bigr)^{2}
 +(1-s)\frac{\beta }{M}m^{x}(\alpha ) -\tilde{m}^{x}(\alpha )m^{x}(\alpha)\Bigr)\Bigr]
\notag \\
&\hphantom{={}}\times
 \prod_{i=1}^{N}\exp [\tilde{m}^{z}(\alpha )\sigma _{i}^{z}(\alpha )
 +\tilde{m}^{x}(\alpha )\sigma _{i}^{x}(\alpha )]
 \langle \sigma _{i}^{z}(\alpha )|\sigma _{i}^{x}(\alpha ) \rangle
 \langle \sigma _{i}^{x}(\alpha )|\sigma _{i}^{z}(\alpha +1 ) \rangle .
\end{align}
We have neglected a few irrelevant constants.
Since the spin product terms have disappeared,
we can perform the summation over all spin configurations independently
at each site.
Then, we obtain
\begin{align}
& Z_{M} = \idotsint \prod_{\alpha =1}^{M}
dm^{z}(\alpha )\,d\tilde{m}^{z}(\alpha )\,
dm^{x}(\alpha )\,d\tilde{m}^{x}(\alpha )
\notag \\
& \times
 \exp \Bigl[N\sum _{\alpha =1}^{M}
 \Bigl(s\lambda \frac{\beta }{M}\bigl(m^{z}(\alpha )\bigr)^{p}
 -\tilde{m}^{z}(\alpha )m^{z}(\alpha )\Bigr)\Bigr]
\notag \\
& \times
 \exp \Bigl[N\sum _{\alpha =1}^{M}
 \Bigl(-s(1-\lambda )\frac{\beta }{M}\bigl(m^{x}(\alpha )\bigr)^{2}
 + (1-s)\frac{\beta }{M}m^{x}(\alpha )
 - \tilde{m}^{x}(\alpha )m^{x}(\alpha)\Bigr)\Bigr]
\notag \\
& \times
 \exp \Bigl[N\ln \Tr \prod_{\alpha =1}^{M}\exp[\tilde{m}^{z}(\alpha )\sigma ^{z}(\alpha )
 + \tilde{m}^{x}(\alpha )\sigma ^{x}(\alpha )]
 \langle \sigma ^{z}(\alpha )|\sigma ^{x}(\alpha ) \rangle
 \langle \sigma ^{x}(\alpha )|\sigma ^{z}(\alpha +1) \rangle\Bigr],\label{eq:Z 2}
\end{align}
where the trace means the summation over the spin variables
$\sigma ^{z}(\alpha ),\sigma ^{x}(\alpha )$ $(\alpha =1,\dotsc ,M)$.

Note that the exponent in Eq.~(\ref{eq:Z 2}) is proportional to $N$.
Thus, the integrals over the variables are evaluated by the saddle point method,
which is to take the maximum value of the integrand as the result of integral
(see, e.g., Appendix~A.1 of \cite{nishimori11:_eleme_phase_transition}).
The saddle point conditions
for $m^{z}(\alpha )$ and $m^{x}(\alpha )$ lead to
\begin{align}
 \tilde{m}^{z}(\alpha ) &= \frac{\beta }{M}ps\lambda \bigl(m^{z}(\alpha )\bigr)^{p-1},\label{eq:sp tilde mz}
\\
 \tilde{m}^{x}(\alpha ) &= \frac{\beta }{M}\{(1-s)-2s(1-\lambda)m^{x}(\alpha )\}.\label{eq:sp tilde mx}
\end{align}

We now use the static approximation, which removes all the $\alpha $ dependence
of the parameters.
We will check the validity of this approximation in Sec.~\ref{sec:results}.
After this approximation, we can easily take trace in Eq.~(\ref{eq:Z 2})
by the converse operation of the Trotter decomposition.
Then, using Eqs.~(\ref{eq:sp tilde mz}) and (\ref{eq:sp tilde mx}),
we finally obtain
\begin{equation}
 Z= \iint dm^{z}\,dm^{x}\, \exp [-N\beta f(\beta ,s,\lambda ;m^{z},m^{x})],
\end{equation}
where $f(\beta ,s,\lambda ;m^{z},m^{x})$ is the pseudo free energy defined as follows:
\begin{align}
& f(\beta ,s,\lambda ;m^{z},m^{x}) =
  (p-1)s\lambda (m^{z})^{p} -s(1-\lambda )(m^{x})^{2}
\notag \\
& -\frac{1}{\beta }\ln 2\cosh \beta
 \sqrt{\bigl\{ps\lambda (m^{z})^{p-1}\bigr\}^{2}
 + \bigl\{1-s-2s(1-\lambda )m^{x}\bigr\}^{2}}.\label{eq:pseudo free energy in beta}
\end{align}
The saddle point equations are thus
\begin{align}
& m^{z} = \frac{ps\lambda (m^{z})^{p-1}}%
  {\sqrt{
  \bigl\{ps\lambda (m^{z})^{p-1}\bigr\}^{2} + \bigl\{1-s-2s(1-\lambda )m^{x}\bigr\}^{2}
  }}
\notag \\
& \times
\tanh \beta \sqrt{
  \bigl\{ps\lambda (m^{z})^{p-1}\bigr\}^{2} + \bigl\{1-s-2s(1-\lambda )m^{x}\bigr\}^{2}},\label{eq:self consistent eq for mz in beta}
\\
&m^{x} = \frac{1-s-2s(1-\lambda )m^{x}}%
  {\sqrt{
  \bigl\{ps\lambda (m^{z})^{p-1}\bigr\}^{2} + \bigl\{1-s-2s(1-\lambda )m^{x}\bigr\}^{2}
  }}
\notag \\
& \times
\tanh \beta \sqrt{
  \bigl\{ps\lambda (m^{z})^{p-1}\bigr\}^{2} + \bigl\{1-s-2s(1-\lambda )m^{x}\bigr\}^{2}}.\label{self consistent eq for mx in beta}
\end{align}

\subsection{\label{subsec:low temp limit}Low temperature limit}
We next derive self-consistent equations in the low temperature limit
to examine quantum phase transitions.
Since the start of the QA process belongs to the paramagnetic phase
and the goal is the ferromagnetic phase,
a quantum phase transition inevitably occurs
in the course of time evolution.

It is useful to consider two possibilities separately depending on whether
the argument of the square root in
Eqs.~(\ref{eq:self consistent eq for mz in beta}) and
(\ref{self consistent eq for mx in beta}) vanishes or not.
We start our discussion from the latter case.

When the square root in Eqs.~(\ref{eq:self consistent eq for mz in beta}) and (\ref{self consistent eq for mx in beta}) assumes a finite value,
the hyperbolic tangent tends to unity
in the $\beta \to \infty $ limit.
Then, we have
\begin{align}
& m^{z} = \frac{ps\lambda (m^{z})^{p-1}}%
  {\sqrt{
  \bigl\{ps\lambda (m^{z})^{p-1}\bigr\}^{2} + \bigl\{1-s-2s(1-\lambda )m^{x}\bigr\}^{2}
  }},\label{eq:self consistent eq for mz}
\\
&m^{x} = \frac{1-s-2s(1-\lambda )m^{x}}%
  {\sqrt{
  \bigl\{ps\lambda (m^{z})^{p-1}\bigr\}^{2} + \bigl\{1-s-2s(1-\lambda )m^{x}\bigr\}^{2}
  }}.\label{eq:self consistent eq for mx}
\end{align}
The pseudo free energy~(\ref{eq:pseudo free energy in beta})
becomes
\begin{align}
& f(s,\lambda ;m^{z},m^{x}) = (p-1)s\lambda (m^{z})^{p} -s(1-\lambda )(m^{x})^{2}
\notag \\
& -\sqrt{\bigl\{ps\lambda (m^{z})^{p-1}\bigr\}^{2}
 + \bigl\{1-s-2s(1-\lambda )m^{x}\bigr\}^{2}}.\label{eq:pseudo free energy trivial}
\end{align}

Equations~(\ref{eq:self consistent eq for mz}) and (\ref{eq:self consistent eq for mx})
have a ferromagnetic (F) solution with $m^{z}>0$ and a quantum paramagnetic (QP)
solution satisfying $m^{z}=0$ and $m^{x}\ne 0$.
Substitution of $m^{z}=0$ into Eq.~(\ref{eq:self consistent eq for mx})
yields
\begin{equation}
m^{x} = \frac{1-s-2s(1-\lambda )m^{x}}{\lvert 1-s-2s(1-\lambda )m^{x}\rvert},\label{eq:QP solution mx}
\end{equation}
i.e., $m^{x}$ can be $\pm 1$.
However, $m^{x}=-1$ is not a proper solution since, with $m^{x}=-1$,
$1-s-2s(1-\lambda )m^{x} = 1-s +2s(1-\lambda )\ge 0$ for $0\le s \le 1$,
$0\le \lambda \le 1$, which leads to $m^{x} =1$ according to Eq.~(\ref{eq:QP solution mx}).
The other possibility $m^{x}=1$ satisfies Eq.~(\ref{eq:QP solution mx})
when $s < 1/(3-2\lambda )$.
Therefore the QP phase can exist in the region $0\le s <1/(3-2\lambda )$,
and its free energy is
\begin{align}
 f_{\text{QP}}(s,\lambda ) = -s\lambda +2s -1,\label{eq:fQP}
\end{align}
which is independent of $p$.

The free energy of the F phase cannot be obtained analytically for general $p$.
However, we can evaluate it in the $p\to \infty$ limit as follows:
In this limit, Eq.~(\ref{eq:self consistent eq for mz}) reads
$m^{z}=0$ or $m^{z} =1$.
The latter solution corresponds to the F phase.
The magnetization in the $x$~direction is zero since Eqs.~(\ref{eq:self consistent eq for mz}) and (\ref{eq:self consistent eq for mx})
satisfy $(m^{z})^{2} + (m^{x})^{2} = 1$.
Substituting the values of magnetization into Eq.~(\ref{eq:pseudo free energy trivial})
and taking the limit $p\to \infty $, we find
\begin{align}
f_{\text{F}}(s,\lambda )\rvert_{p\to \infty} =& -s\lambda .\label{eq:fF}
\end{align}

Let us next consider the case, where the argument of square root
in Eqs.~(\ref{eq:self consistent eq for mz in beta}) and
(\ref{self consistent eq for mx in beta}) vanishes.
We then assume that $m^{z}$ and $m^{x}$ tend to the following values
as $\beta \to \infty $:
\begin{equation}
  m^{z} \to  0,\quad
  m^{x} \to \frac{1-s}{2s(1-\lambda )}\label{eq:limit of mz and mx}
\end{equation}
such that the argument of hyperbolic tangent approaches a finite constant:
\begin{equation}
 \beta \sqrt{\bigl\{ps\lambda (m^{z})^{p-1}\bigr\}^{2}
 + \bigl\{1-s-2s(1-\lambda )m^{x}\bigr\}^{2}} \to c.\label{eq:assumption of argument}
\end{equation}
In order to find a non-trivial solution,
it is also necessary to assume the following relation:
\begin{equation}
 \frac{ps\lambda (m^{z})^{p-1}}{1-s-2s(1-\lambda )m^{x}} \to 0.
\end{equation}
Under these assumptions, Eqs.~(\ref{eq:self consistent eq for mz in beta}) and
(\ref{self consistent eq for mx in beta})
read $m^{z} = 0$ and $m^{x} = \tanh c$.
These equations satisfy the condition~(\ref{eq:limit of mz and mx})
if we choose $c$ such that $\tanh c = (1-s)/2s(1-\lambda )$.

Unless $s=1$, the magnetizations~(\ref{eq:limit of mz and mx}) satisfy
the condition of QP solution.
we then call this phase QP2 in order to distinguish from the QP phase
described before.
The free energy of QP2 phase is obtained in
the limit~(\ref{eq:limit of mz and mx}) and $\beta \to \infty$
under the assumption~(\ref{eq:assumption of argument}):
\begin{equation}
 f_{\text{QP2}}(s,\lambda ) = -\frac{(1-s)^{2}}{4s(1-\lambda )}.\label{eq:free energy for QP2}
\end{equation}
The domain of applicability of the free energy~(\ref{eq:free energy for QP2})
is restricted by $1/(3-2\lambda ) \le s < 1$
since $\left| (1-s)/2s(1-\lambda )\right| = \left| \tanh c\right| \le 1$
and $s \ne 1$.
This region of $s$ will be called the QP2 domain hereafter.

\subsection{\label{subsec:lambda zero}%
Phase transition on the line $\lambda = 0$}
Although we cannot solve explicitly the self-consistent equations
for general values of the parameters,
it is possible to solve them in the case of $\lambda = 0$.
In the following discussion, we show that a second-order phase transition
occurs at the point $(s,\lambda )=(1/3,0)$ for any value of $p$.
This result is independent of the target Hamiltonian.

We start from the analysis of the phase diagram on this line.
In this case, Eq.~(\ref{eq:self consistent eq for mz}) reduces to $m^{z} = 0$.
It thus suffices to consider the QP and QP2 phases.
These phases do not have a common domain of definition:
The QP and QP2 phases are defined on the region
$0\le s <1/3$ and $1/3\le s < 1$, respectively.
Thus, the former (latter) range is the QP (QP2) phase,
and a phase transition exists at $s=1/3$.

The magnetization in the $x$~direction of
the QP2 phase~(\ref{eq:limit of mz and mx})
is identical to that of the QP phase at the transition point.
This means that the transition is of second order.
The free energy $f_{\text{QP}}(s,0)$, of course, smoothly connects to
$f_{\text{QP2}}(s,0)$ at the point~\cite{footnote:phase_trans_QP2_QP}.

Though the existence of a second-order transition does not hamper
the efficiency of QA~\cite{schuetzhold06:_adiab_quant_algor_quant_phase_trans},
paths of QA must not follow this line
because quantum fluctuations completely disappear on this line:
The total Hamiltonian $\hat{H}(s,0)$ is diagonalized in the $x$~basis.
Quantum state transitions do not occur,
and the system does not perform quantum annealing processes.

In the following section, we will show that paths exist to avoid first-order
transition in the region $\lambda >0$ at least for $5\le p \le 21$,
and possibly for any large but finite $p$.

\section{\label{sec:results}Numerical results}

\subsection{\label{subsec:phase diagram}Phase diagram}
Let us next analyze numerically the phase diagram
on the $s$-$\lambda $ plane for finite
values of $p$. We construct the phase diagram as follows.
We first solve numerically the self-consistent equations~(\ref{eq:self consistent eq for mz})
and (\ref{eq:self consistent eq for mx}) for a given value of $p$
and at a point $(s,\lambda )$ in the phase diagram and then evaluate
the corresponding free energy.
By comparing all possible solutions and their free energies
including $f_{\text{QP2}}$,
we identify the stable solution having the smallest value of the free energy.

\begin{figure}[t]
  \centering
  \includegraphics[width=0.7\hsize, clip]{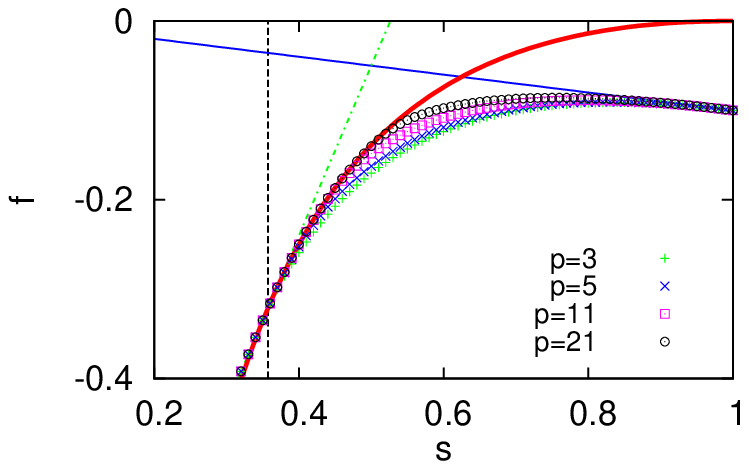}\\
  \includegraphics[width=0.7\hsize, clip]{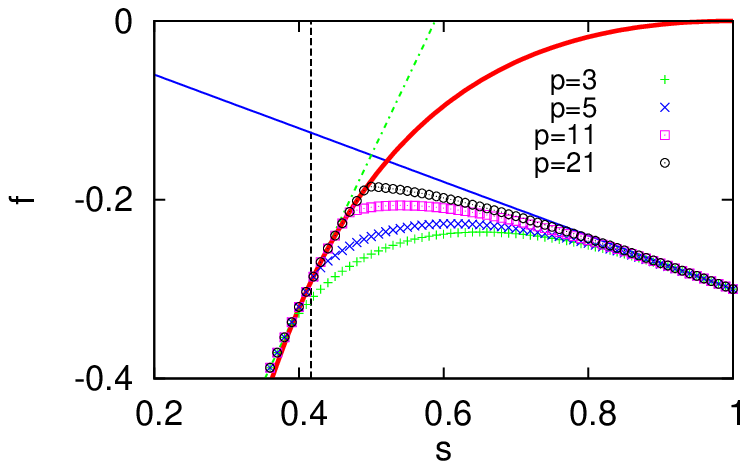}\\
  \caption{\label{fig:free energies}
  Free energy vs.\ $s$ for some values of $p$.
  The parameter $\lambda $ is 0.1 (top), and 0.3 (bottom).
  The dash-dotted line in light green represents the free energy of the QP phase, Eq.~(\ref{eq:fQP}),
  the thin solid line in blue is for the F phase, Eq.~(\ref{eq:fF}),
  and the thick solid line in red for the QP2 phase,
  Eq.~(\ref{eq:free energy for QP2}).
  The vertical dashed line denotes the lower
  limit of the QP2 domain ($s=1/(3-2\lambda )$).
  Although it is difficult to discern in the present scale,
  all the data for finite $p$ we studied have lower values
  than that of $f_{\text{QP2}}$ in the QP2 domain.}
 \end{figure}

It is useful to show the dependence of the free energy
on $s$ for some values of $p$ and $\lambda $
as in Fig.~\ref{fig:free energies}.
We have confirmed numerically that the free energy lies below $f_{\text{QP2}}$
in the QP2 domain, and the QP2 phase is completely suppressed by the other phases.
This system thus undergoes a quantum phase transition
from the QP phase for small $s$ to the F phase for large $s$.

\begin{figure}[t]
  \centering
  \includegraphics[width=0.7\hsize, clip]{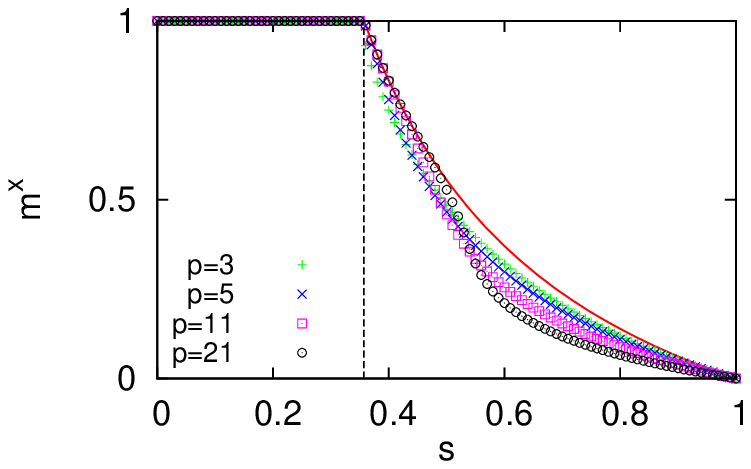}\\
  \includegraphics[width=0.7\hsize, clip]{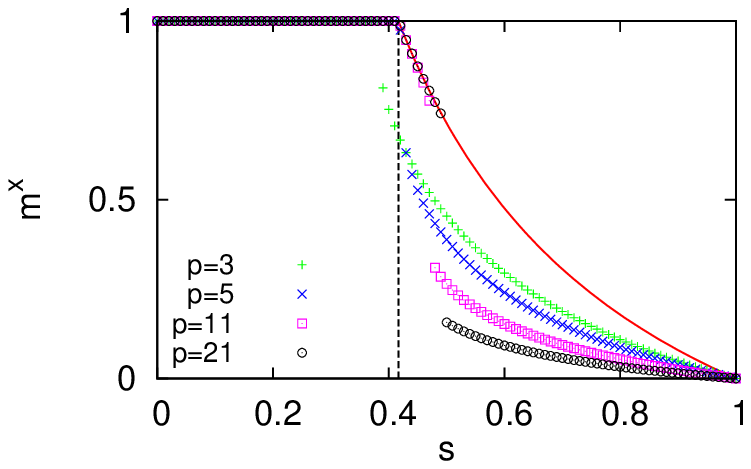}\\
  \caption{\label{fig:magnetizations}%
  Magnetization $m^{x}$ corresponding to Fig.~\ref{fig:free energies}.
  The solid line represents the $x$~component of magnetization
  of the QP2 phase~(\ref{eq:limit of mz and mx}),
  and the vertical dashed lines are the same as those in Fig.~\ref{fig:free energies}.
  For $\lambda = 0.1$ (top panel) and $p \ge 5$, a second-order transition occurs
  at the boundary of the QP2 domain; The magnetization decreases continuously
  from unity to zero. In contrast, the magnetization for $\lambda =0.3$ (bottom panel)
  has a jump.}
 \end{figure}

To determine whether the transition is first order or second order,
we show the behavior of the magnetization $m^{x}$ in
Fig.~\ref{fig:magnetizations}.
The parameters of the figure correspond to those in
Fig.~\ref{fig:free energies}.
When $\lambda = 0.1$,
the magnetization $m^{x}$ for $p=3$ has a small jump at $s=0.3544(1)$,
and $m^{x}$ for $p \ge 5$ decreases continuously from unity to our numerical precision.
This means that $m^{z}$ for $p \ge 5$ increases continuously
from zero to a finite value.
Therefore a second-order transition
occurs for $p \ge 5$ at $\lambda = 0.1$.
The same is true for $\lambda = 0.3$ in the sense that there exists a second-order
transition at the boundary of the QP2 phase for $p\ge 5$.

A remarkable fact is that the magnetization
for some parameters (e.g., $\lambda =0.3$, $p=11$)
in Fig.~\ref{fig:magnetizations}
jumps within the F phase.
This discontinuity results in an exponential decrease of the energy gap
as $N$ increases.
There exists a first-order transition within the F phase.
However, this unusual behavior disappears for smaller values of $\lambda $
for any finite $p$, excluding $p=3$, that we checked.
Thus for smaller $\lambda $, only a second-order transition takes place
as we increase $s$ from zero to a value close to unity.

The resulting phase diagrams are shown in Fig.~\ref{fig:phase diagram for p=3,5,11}
for $p=3,5$, and $11$.
We see that a boundary of second-order transition exists for small $\lambda $
and $p\ge 5$.
It is observed that one can reach the F phase from the QP phase by choosing
a path that avoids a first-order transition
as long as the first-order F-F boundary does not reach the $\lambda = 0$ axis,
which happens probably only in the limit $p\to \infty $ as we shall see below.

 \begin{figure*}[t]
  \centering
  \includegraphics[width=\hsize, clip]{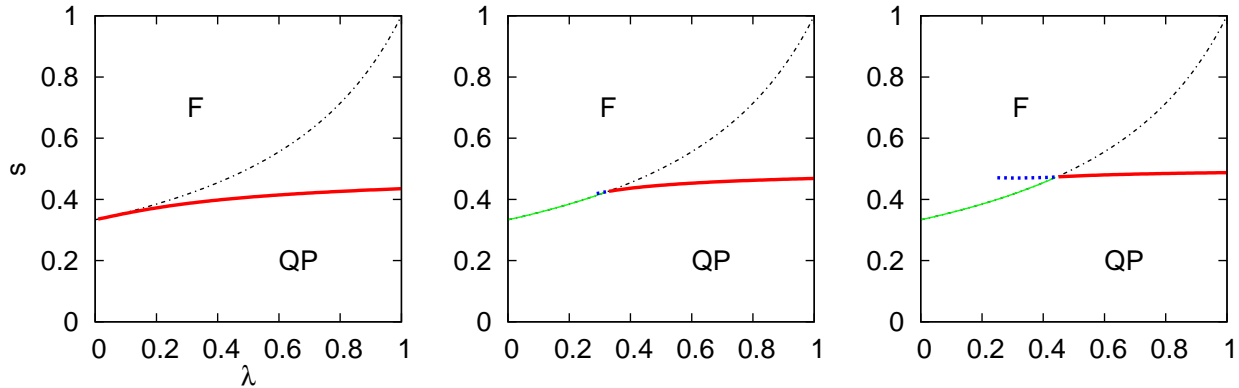}
  \caption{\label{fig:phase diagram for p=3,5,11}%
  Phase diagrams on the $s$-$\lambda $ plane for $p=3$ (left), $p=5$ (middle),
  and $p=11$ (right).
  The dash-dotted line represents the boundary of the QP2 domain
  ($s=1/(3-2\lambda )$), where a transition takes place between the QP and F phases.
  For large $\lambda $, the QP and F phases are separated
  by the horizontal phase boundary (QP-F boundary).
  The thick solid line in red represents the first-order transition,
  and the thin solid line in light green is for the second-order transition.
  For $p=5$ and 11, the magnetization jumps on the dashed line in blue (F-F boundary)
  within the F phase~\cite{footnote:TAF}.}
 \end{figure*}

\subsection{\label{subsec:energy gap}Energy gap}
We next study the behavior of the energy gap across the phase transitions
found in the previous section.

To calculate the energy gap for large $N$,
we adopt the method used in~\cite{Jorg2010energy}.
The Hamiltonian under consideration is expressed by
the components of total spin operator $\hat{S}^{x,z}$,
thus commuting with the total spin $\hat{\bm{S}}$.
Since the total angular momentum is conserved during the time evolution,
we have to pay attention only to the subspace
that has the maximum angular momentum $S=N/2$.
The dimension of this subspace is $N+1$,
which greatly enhances the possible system size to $N\sim 100$.

\begin{figure}[t]
  \centering
  \includegraphics[width=0.7\hsize, clip]{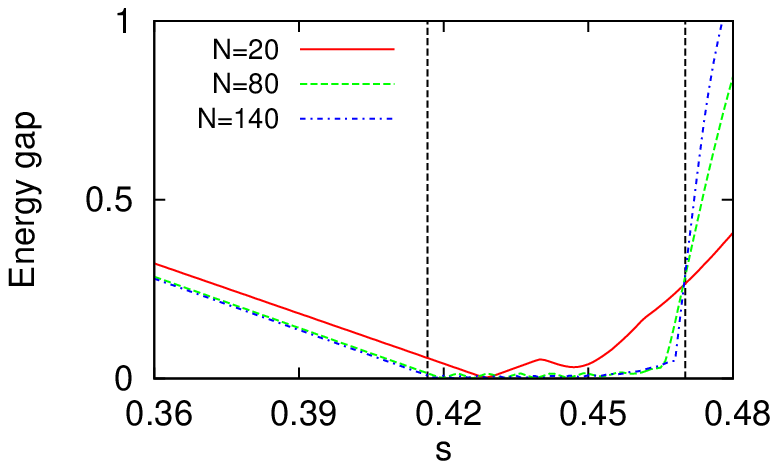}\\
  \includegraphics[width=0.7\hsize, clip]{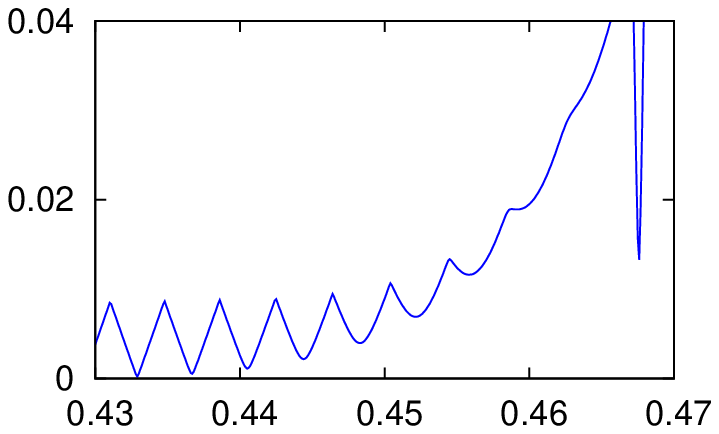}
  \caption{\label{fig:enegy gap p=11 lam=0.3}%
  Top panel: Energy gap vs.\ $s$ for $p=11$ and $\lambda =0.3$.
  The vertical dashed lines represent the boundary of the QP2 domain
  at $s\simeq 0.4167$ and the F-F boundary at $s\simeq 0.4701$.
  The bottom panel is the enlarged view of the top panel for $N=140$.}
 \end{figure}

It is useful to first verify the validity of the static approximation.
Figure~\ref{fig:enegy gap p=11 lam=0.3}
shows a representative energy gap
with a second-order phase transition:
As one sees in the enlarged view shown in the bottom panel,
the gap shows wiggly behavior for a finite range.
The wiggly behavior starts at $s\simeq 0.4184$ for $\lambda = 0.3$,
which corresponds to the left end of the QP2 domain and also
to the second-order transition point between the QP and F phases.
The same behavior terminates at $s\simeq 0.4676$ for $\lambda = 0.3$,
corresponding to the first-order F-F boundary.
These two transition points evaluated analytically
using the static approximation, Eqs.~(\ref{eq:self consistent eq for mz})
and (\ref{eq:self consistent eq for mx}),
are shown in dashed vertical lines in Fig.~\ref{fig:enegy gap p=11 lam=0.3}
and agree fairly satisfactorily with the numerical results,
as $N$ increases, for the interval where the gap is very small.

\begin{figure}[t]
  \centering
  \includegraphics[width=0.7\hsize, clip]{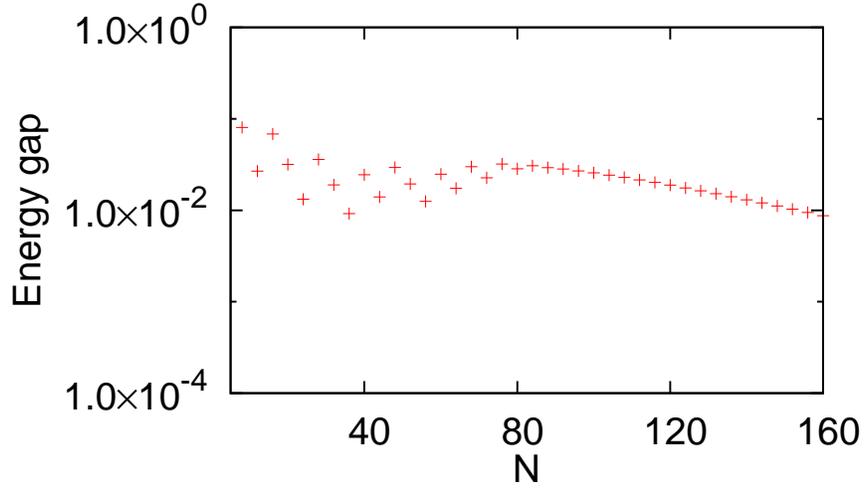}
  \caption{\label{fig:rightmost gap}%
  The rightmost local minimum of the energy gap as a function of $N$
  for $p=11$ and $\lambda =0.3$ on a semi-log scale.
  The gap closes exponentially with $N$.}
 \end{figure}

The rightmost local minimum of the energy gap in Fig.~\ref{fig:enegy gap p=11 lam=0.3}
behaves differently from other local minima and decays exponentially as $N$ increases
as shown in Fig.~\ref{fig:rightmost gap}.
This is expected from the jump in the magnetization shown
in Fig.~\ref{fig:magnetizations} because a jump implies a first-order transition
though the system is ferromagnetic in both sides of the transition point.
Although this is not the global minimum, it will affect the efficiency of QA
for much larger systems where the rightmost one will become the global minimum since
the other local minima decay only polynomially as shown below.

\begin{figure}[t]
  \centering
  \includegraphics[width=0.7\hsize, clip]{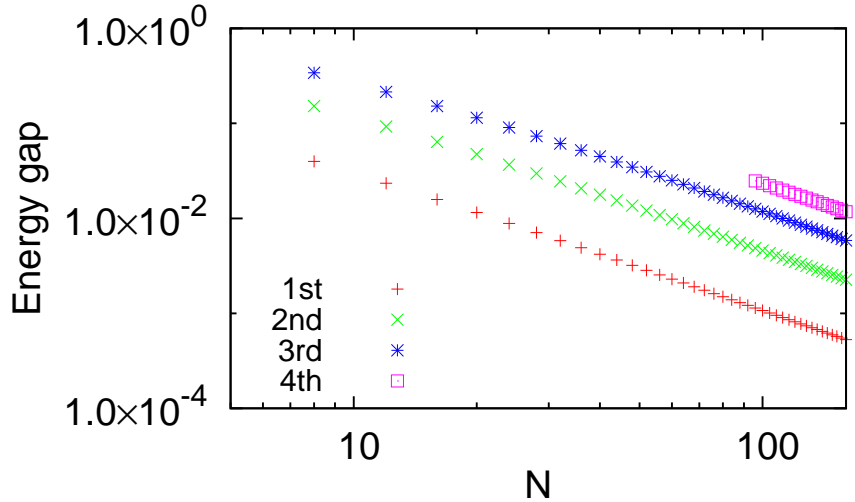}\\
  \caption{\label{fig:deltamins p=5 lam=0.1}%
  Energy gap vs.\ $N$ at local minima for $p=5,\lambda =0.1$
  on a log-log scale.
  We number the minima from left to right.
  No gaps vanish exponentially up to the size studied here.}
 \end{figure}

\begin{figure}[t]
  \centering
  \includegraphics[width=0.7\hsize, clip]{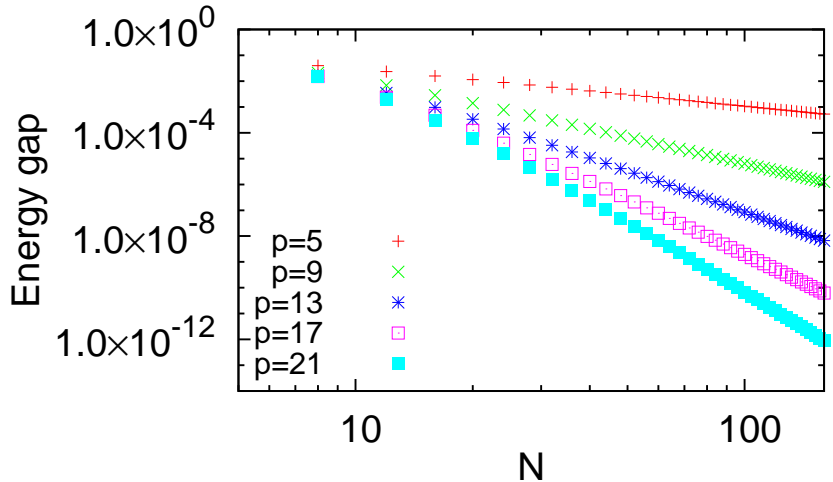}\\
  \caption{\label{fig:deltamin p=5 to 21 lam=0.1}%
  Minimum gap vs.\ $N$ for some values of $p$ on a log-log scale for $\lambda = 0.1$.
  Each data scales polynomially.}
 \end{figure}

Figure~\ref{fig:deltamins p=5 lam=0.1} shows the size dependence of local minima
of the energy gap for $p=5$ and $\lambda =0.1$.
All minima shown here decay polynomially.
In Fig.~\ref{fig:deltamin p=5 to 21 lam=0.1} the global minimum of energy gap
for selected $p$ is depicted as a function of $N$ at $\lambda =0.1$.
For any value of $p$, the gap closes polynomially at least up to
the system size we studied, $N=160$.

The above results suggest that first-order transitions will be able to be avoided
if we choose a path around $\lambda =0.1$ when we reach the F phase
from the QP phase by increasing $s$ as long as $p$ is not too small
and not too large, $5\le p \le 21$.
It is then interesting to see what happens in the limit of large $p$.

\section{\label{sec:phase diagram for p infinity}%
Phase diagram in the infinite-$p$ limit}
The ferromagnetic $p$-spin model reduces to the Grover problem
in the $p\to \infty $ limit~\cite{Jorg2010energy, grover97:_quant_mechan_helps_searc_needl_hayst}.
The goal of the Grover problem is to find the desired item
in an unsorted database containing $2^{N}$ items.
Whereas classical algorithms require a time of $O(2^{N})$ to find the desired item,
the quantum algorithm, called the Grover algorithm, costs only a time of $O(2^{N/2})$,
quadratic speed-up
though the time complexity still scales exponentially.

Farhi \textit{et al.}\ have proposed a QA version of
the Grover algorithm~\cite{farhi:_quant_comput_adiab_evolut}, which adopts the transverse field as a driver Hamiltonian.
Unfortunately, the time complexity is the same as that of classical algorithms.
However, Roland and Cerf have improved the efficiency of QA by adjusting
the evolution rate $s(t)$, then reproducing the quadratic speed-up,
and they have proved that their algorithm is optimal~\cite{roland02:_quant_searc_local_adiab_evolut}.
This result indicates that our approach cannot avoid jumps of magnetization
in the $p \to \infty $ limit.
It is therefore interesting to study how this difficulty appears in our method.

To this end, it is instructive to study the behavior of the free energy
and magnetization for large but finite values of $p$.
The free energy in Fig.~\ref{fig:free energies} is seen to approach
the asymptotic values in Eqs.~(\ref{eq:fF}) and (\ref{eq:free energy for QP2}) from below.
Hence, the QP2 phase does not appear for any finite $p$.
From Fig.~\ref{fig:magnetizations}, we observe that the magnetization
in the $x$~direction is close to the QP2 phase magnetization~(\ref{eq:limit of mz and mx}), shown in red solid lines,
in the region where the free energy approaches $f_{\text{QP2}}$.
The magnetization in the $z$~direction is
\begin{align}
m^{z} = \sqrt{1-\left(\frac{1-s}{2s(1-\lambda )}\right)^{2}} \not\equiv 0
\label{eq:mpq2}
\end{align}
since the QP2 phase does not appear.

\begin{figure}[t]
  \centering
  \includegraphics[clip]{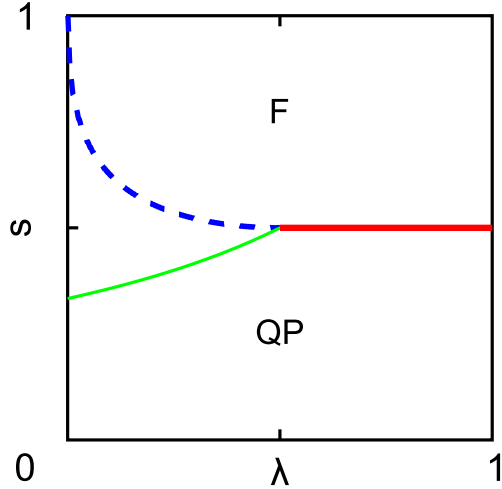}\\
  \caption{\label{fig:phase diagram for infinite p}%
  Phase diagram in the limit $p\to \infty$.
  Three lines represent the same phase boundary
  as those in Fig.~\ref{fig:phase diagram for p=3,5,11}.
  The QP phase has the magnetization $m^{z}=0$.
  The F phase above the F-F boundary, shown dashed in blue,
  has the magnetization $m^{z}=1$ and
  the phase below the F-F boundary has $0<m^{z}<1$.}
 \end{figure}

We extrapolate these results to the case of $p\to \infty $.
That is, while the free energies are described by Eqs.~(\ref{eq:fQP}),
(\ref{eq:fF}), and (\ref{eq:free energy for QP2}),
the magnetization in the QP2 phase is given by Eq.~(\ref{eq:mpq2}).
To be precise, this is not the QP2 phase
since the magnetization in the $z$~direction is nonzero.
With a caution on the domain of QP and QP2 in mind, we compare the values
of the free energy of the three phases and obtain the phase diagram
as in Fig.~\ref{fig:phase diagram for infinite p}.
The F phase and the QP phase are separated by a horizontal phase boundary.
The boundary of second-order transition is given by
$s=1/(3-2\lambda )\ (\lambda \le 1/2)$, and
the first-order F-QP transition boundary is $s=1/2\ (\lambda >1/2)$.
Solving $f_{\text{F}}\rvert _{p\to \infty } = f_{\text{QP2}}$, we get the F-F boundary
as
\begin{align}
s =& \frac{1-2\sqrt{\lambda -\lambda ^{2}}}{(2\lambda -1)^{2}}.
\end{align}
The figure shows that an abrupt change of magnetization, a first-order transition,
is inevitable in the limit $p\to \infty $.

\section{\label{sec:conclusion}Conclusion}
In the present paper, we have introduced a new approach to QA using
antiferromagnetic quantum fluctuations.
This approach adopts two types of the driver Hamiltonian,
the transverse-field term $\hat{V}_{\text{TF}}$ and
the transverse antiferromagnetic two-body interaction term $\hat{V}_{\text{AFF}}$
in Eq.~(\ref{eq:Vaff}).

We have applied this method to the ferromagnetic
$p$-spin model, which was considered to be hard to find the ground state
for $p>2$ with a simple QA~\cite{Jorg2010energy}.
We have evaluated the efficiency from the phase diagram
and the minimum values of the energy gap.
Numerical calculations have shown that the phase boundary of second-order transition
appears for $p \ge 5$.
However, the magnetization in the F phase jumps for large $p$.
Although the boundary at which the magnetization in the F phase jumps extends
as $p$ increases, we have confirmed numerically that there remains a region
where the magnetization changes continuously at least for $5\le p \le 21$
when $\lambda = 0.1$.
This indicates that QA can solve the problem efficiently in this case.
In fact, we have calculated the minimum gap up to $N=160$
and have confirmed that it vanishes polynomially
on the second-order phase boundary.
Thus, QA with antiferromagnetic fluctuations
is an efficient algorithm at least for $5\le p \le 21$.
We expect to be able to avoid the difficulty of exponential complexity
for larger value of $p$ as long as it is finite.
It is an interesting problem to study if the present method
improves the efficiency of QA for other systems.


\begin{thebibliography}{99}
\bibitem{footnote:phase_trans_QP2_QP}%
There is no order parameter to characterize this phase transition in the sense
the order parameter is finite in one phase and zero in the other.
We nevertheless call it a phase transition because the energy (free energy) is
singular at $s=1/3$, $f''_{\text{QP}}(s=1/3-0,0) \ne f''_{\text{QP2}}(s=1/3+0,0)$.
A similar phenomenon is observed within the ferromagnetic phase,
a phase transition from a ferromagnetic phase to another ferromagnetic phase,
as will be shown in Sec.~\ref{sec:results}.
\bibitem{footnote:TAF}%
It is to be noted that the special point $(s,\lambda ) = (1,0)$ has a highly
degenerate ground state~\cite{chandra:_quant_phase_trans_in_disor_long-range},
among which only the state with the maximum total spin $S=N/2$ is relevant to
our purpose: Quantum annealing starts from $s=0$ with $S=N/2$ and preserves
the total spin following the Schr\"{o}dinger equation~(\ref{eq:schrodinger equation}) with $[\hat{H}(t),\hat{\bm{S}}^{2}]=0$.
\end{thebibliography}

\end{document}